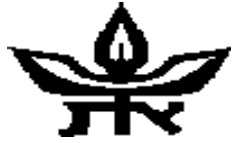

TEL-AVIV UNIVERSITY

The Raymond and Beverly Sackler

Faculty of Exact Sciences

School of Computer Sciences

# CDTP
# Chain Distributed Transfer Protocol

Thesis submitted in partial fulfillment of the requirements for the degree of
Master of Science

by
## Shmuel Vagner

Prepared under the supervision of
*Prof. **Yehuda Afek***

December 2002

## Acknowledgements

First of all I would like to thank Prof. Yehuda Afek for his excellent guidance and knowledge and foremost for the patience and time invested in helping me with this endeavour when his time was dear.
I thank Mr. Sharon Sher from RADWIN Ltd. for providing me with equipment to perform the measurements in this work.
Most of all thanks to my wife who made real sacrifices to enable me to take the time to do this work, without her love and understanding this work would not have been possible.

## Abstract

The rapid growth of the internet in general and of bandwidth capacity at internet clients in particular poses increasing computation and bandwidth demands on internet servers. Internet access technologies like ADSL [DSL], Cable Modem and Wireless modem allow internet clients to access the internet with orders of magnitude more bandwidth than using traditional modems. We present CDTP a distributed transfer protocol that allows clients to cooperate and therefore remove the strain from the internet server thus achieving much better performance than traditional transfer protocols (e.g. FTP [FTP]). The CDTP server and client tools are presented also as well as results of experiments. Finally a bandwidth measurement technique is presented. CDTP tools use this technique to differentiate between slow and fast clients.

# Table of Contents

## Table of Figures

# Introduction

As the internet becomes a standard source of information for increasing number of users, access technologies emerge that allow users faster access to the internet. Technologies like ADSL, Cable modem and Wireless modem allow users access rates of several Megabits per second instead of the 56Kbps allowed by the traditional modems. These access techniques and the increasing number of internet users pose increasing bandwidth and computational demands on internet servers since they have to support both increasing numbers of clients and to be able to supply data at much higher rate to each of the clients. Bandwidth and computational power demands of traditional transfer protocols like FTP grow linearly with the number of users and their bandwidth capacity.
Various techniques were developed to solve this problem: Server farms with load balancers enable distributing the load between several physical servers at the site. Such solutions are more fault tolerant than just using a single server but the required bandwidth and the total computational power is still proportional to the number of users and their access speeds. Mirror site arrays like Akamai [AKA] are another way to solve the problem, this solution has the robustness advantage of a load balanced server farm and in addition it allows clients to be served by nearby servers, still from the point of view of the provider the amount of computing power and bandwidth that has to be bought is proportional to the number of users and their speeds.

Another way is to use distributed transfer protocols where clients pass data to other clients. Napster [NAP] and Gnutella [GNU] are examples for such techniques, in both cases clients are expected to store data that they download and to be ready to serve it to other clients that need the data. The problem here is that internet clients that participate in such a network may delete the downloaded data at any time, they are not even guaranteed to be constantly connected to the internet, these facts restrict the usability of such solutions to well known files like music or movie files since 1: The number of such different files is limited and 2: clients tend to store such files for long periods of time.
In this work a new type of distributed protocol is presented it is called CDTP and like Napster and Gnutella it makes internet clients act as servers and transmit data to each other thus relaxing bandwidth demands from the server. Unlike Napster and Gnutella CDTP does not require clients to be constantly logged on to the internet or to store downloaded files for long periods of time, in fact clients do not have to store downloaded files at all, since the only time a client has to serve other clients is when it is downloading files itself. The above relaxation of requirements is based on the observation that as long as clients do not access the server simultaneously the server has no trouble serving them. When a large number clients access a server at the same time (Like in the case when a company like Microsoft releases a new patch or when a record company releases a new single of a known rock band), then the server may have trouble to service all the clients unless enough processing power and bandwidth was prepared in advance. When such situation arises CDTP clients start cooperating and pass the requested data to each other and in this way reduce the load from the server.

To accomplish the above task efficiently it is important that clients will be served by clients with close enough bandwidth capacity, otherwise fast clients will be slowed down by slow clients. A packet pair based bandwidth comparison method is presented, this method measures the capacity of the bottleneck link between a source-destination pair, and it is optimised to the CDTP requirements.

The rest of this paper is structured as follows: Chapter 1 presents existing distributed transfer protocols, chapter 2 presents CDTP, chapter 3 presents existing bandwidth measurement techniques, and the bandwidth measurement technique used by CDTP. chapter 4 describes CDTP in detail, chapter 5 presents results of experiments with CDTP and finally chapter 6 concludes and suggests future directions for research.
The exact CDTP specification resides in appendix A.
A proof of the packet pair property under the model that is defined in this paper is given in appendix B.

# 1. Distributed transfer protocols

## 1.1. Traditional transfer protocols

The traditional and most widely used file transfer and sharing technology is the FTP protocol. In FTP there is a server that holds all the files and clients that can access the server and perform actions for querying the server's contents and getting the files.
The disadvantage of FTP is that it is centralized and in order to serve a large number of clients the server should reside on a powerful computer and have a lot of available bandwidth.

Another related traditional technology is IP Multicast [COM]. IP Multicast is not a parallel technology to FTP (Layer 3 in the OSI Model as opposed to Layer 7), IP Multicast enables hosts to form groups where each member in the group can send IP packets that will reach all the groups members.
This could be a basis for an efficient transfer protocol because once a group is formed the server can send packets to all of the group and thus conserve bandwidth and resources.
The main disadvantage with using IP Multicast for the above purpose is that all the clients should request the file simultaneously, otherwise the server should resend the data and nothing is gained.

## 1.2. Distributed transfer protocols

In recent years file sharing technologies have emerged, these technologies form a virtual network (either centralized or decentralized) that enable users over the internet to share files and this way eliminating the need of a single server to hold all the files.

Napster is an example for a centralized file sharing technology. In Napster a single server holds an index of all the available files and their locations over the internet, when a client wishes to download a file, it access the server with a request and the server directs him to another client that stores the file.

Gnutella is an example for a decentralized file sharing technology. In Gnutella there is no server, instead the hosts in the network that called servants (because they act as both clients and servers) exchange information between themselves about the whereabouts of files in the network using a routing like protocol.

Both file sharing techniques are great for exchanging music and video files between peers over the internet but are less suitable for servers. In Gnutella there is a performance problem because there is no centralized server and therefore clients has to search for the files they need all over the internet. In both Gnutella and Napster files should be kept by clients even after they are no longer needed, which can be reasonable for music files, but unreasonable for downloaded software patches. Also both technologies are not suitable for files that change frequently.

CDTP is a breed between the traditional data transfer methods and the new file sharing methods. It's functionality is similar to FTP and it's implementation is similar to Napster, a major difference between CDTP and Napster is that files are stored in clients only during the time they are being received thus eliminating the two limitations described above.

## 2. CDTP at a glance

In the CDTP network there is a single server and multiple clients. The role of the server is to store and transmit files to the clients and to coordinate between clients that serve each other as will be described below. The clients request files from the server and while the file is being received they feed it to other clients that need the same file.

To service multiple clients that request the same file the server creates chains of clients that service each other. That is the server feeds the file to the first client in the chain which in turn feeds it to the second and so forth. Each chain is identified by the bandwidth between the server and the clients in the chain. See Figure 2 – CDTP Chains.

When a new client requests a file the server first finds an appropriate chain to attach the client to. Essentially the server looks for a chain whose bandwidth is similar to the new client bandwidth and whose end node has not progressed too far in the reception.

The chain is selected and the client is serviced in the following way:
- The server measures the bandwidth between itself and the client.
- The server selects a chain whose bandwidth is compatible to the bandwidth that was measured in the previous step and whose last member has not received more than a predefined fraction of the file. If no such chain is found then the new client starts a new chain.
- The server directs the client to the last client in the selected chain.
- The client requests the file from the last client in the selected chain.
- The last client in the selected chain measures the bandwidth between itself and the new client and if it is no less than the selected chain bandwidth then it begins feeding the new client with the requested file.

If no compatible chain was found then a new chain is formed and the client is fed directly by the server.

The server constantly keeps track on the progress of individual clients by receiving progress reports from them, this way when a chain is broken because one of the clients stopped servicing the client that after him in the chain, the client that stopped receiving data contacts the server with a retransmission request and the server is responsible for finding a new chain for the client with similar amount of progress. If no such chain is found then the server services the client directly by forming a new chain and supplying the rest of the file.

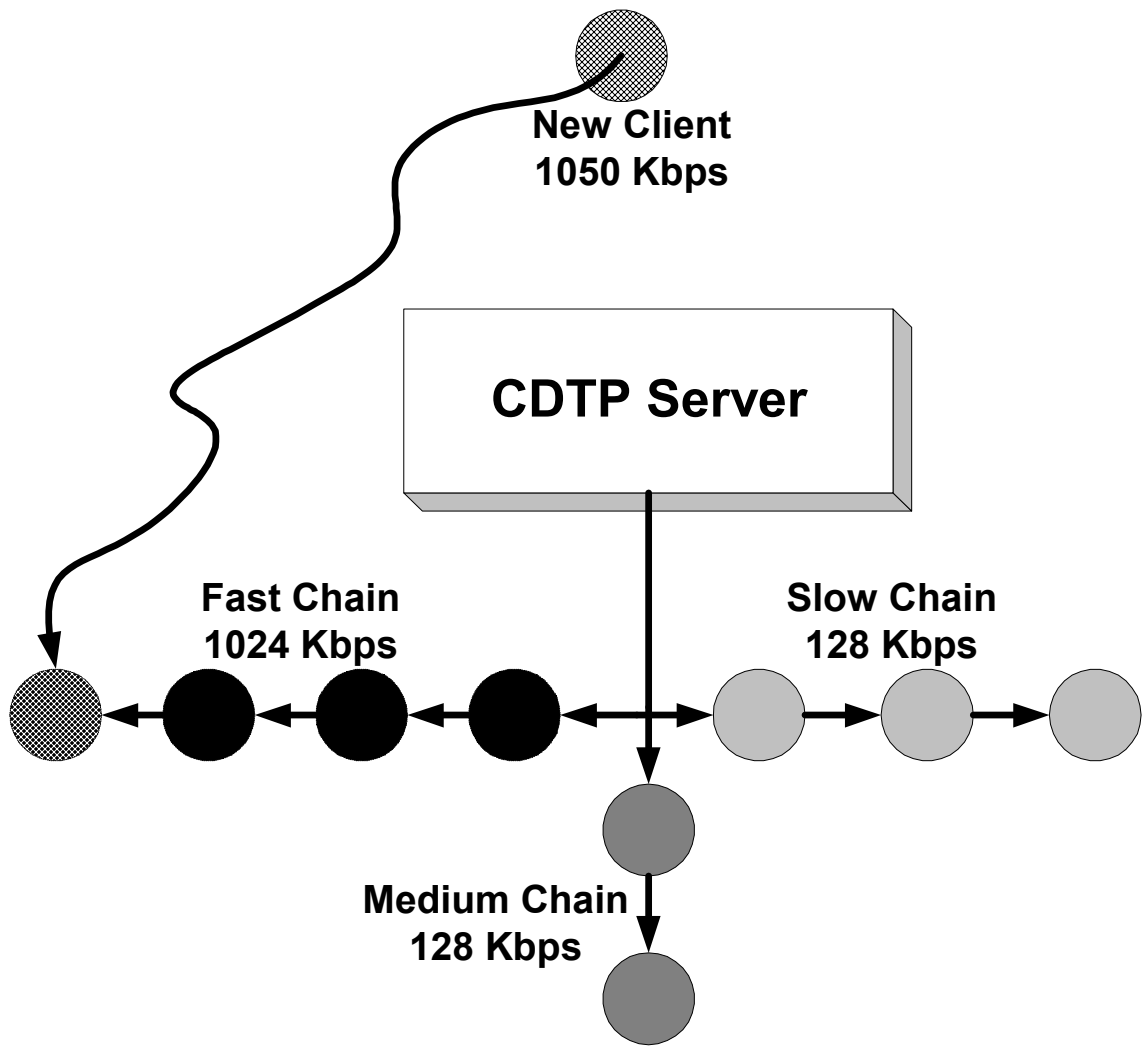

**Figure 1 - Adding a New Client**

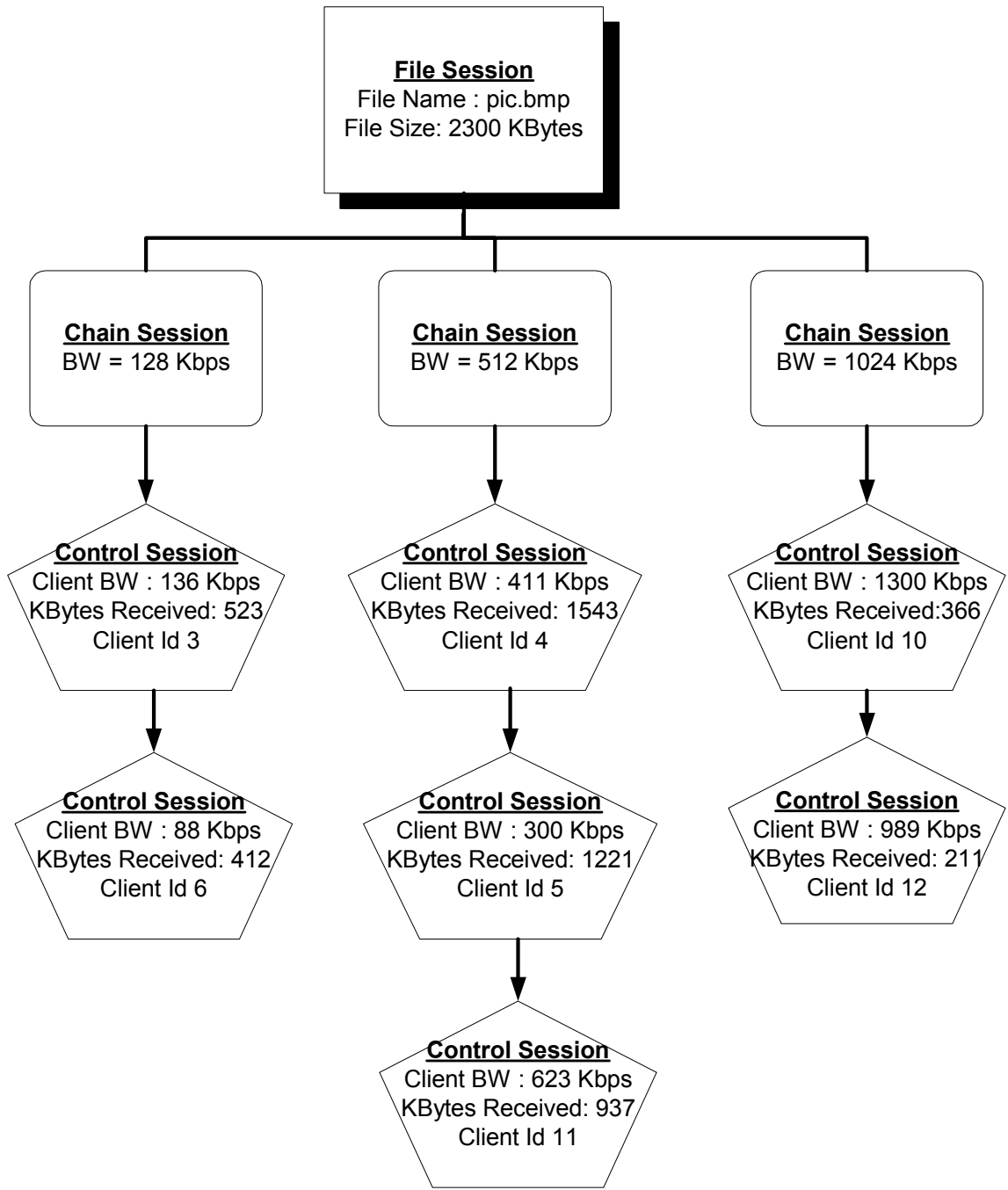

**Figure 2 – CDTP Chains: Data Structures at the Server**

# 3. Bandwidth measurement

In this section a new bandwidth measurement method is presented. It combines two existing techniques, the packet pair technique ( described in 3.2.1) and delay based techniques (described in 3.2.2), the new technique is presented in 3.3.

### *3.1. The Model*

A typical network model is comprised of network hosts that generate and receive traffic, routers that transfer the network traffic and links that connect the above nodes.

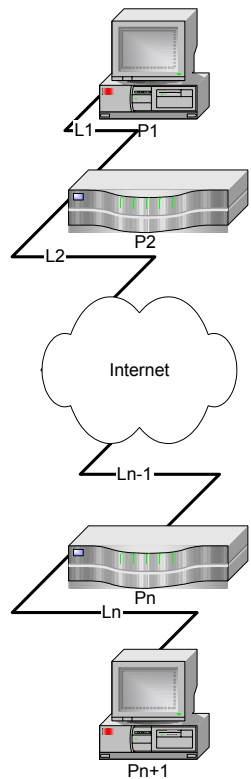

**Figure 3 – Internet Path Model**

The delay of a packet sent between hosts is comprised of the time it takes to send the packet over all of the links plus the sum of latencies of the links plus the sum of queuing delays at the hosts and routers due to processing time and cross traffic. Cross traffic delay is defined as the time a packet is delayed inside a node while waiting for the node to transmit other packets over the next link along it's path .

Denote S as the size of the packet, $b_1,..,b_n$ the link capacities along the path, $l_1,..,l_n$ the link latencies and $q_1,..,q_n$ the queuing delays at nodes along it's path (note then $q_{n+1}$ equals 0). We then get that the time it takes for the packet to reach it's destination is:

$$T = \sum_{i=1}^{i=n} \frac{S}{b_i} + l_i + q_i \qquad (1)$$

Two quantities are interesting when measuring bandwidth of a network path:
The path capacity bandwidth (PCB) which is defined as $Min_{i=1,...,n}(b_i)$ and the path available bandwidth (PAB) which is defined the as $Min_{i=1,...,n}(b_i - c_i)$ where $c_i$ is the cross traffic capacity over the nodes i and i+1. In this work the former is investigated.

### *3.2. Existing measurement techniques*

Two categories of BW measurement exist. One measures PCB (see section 3.2.1) and is based on the packet pair property of network paths that was first identified by [JAC1] . The other measures all the link capacities along a network path (see section 3.2.2) and is based on comparing delays to different nodes along the path. This technique was first introduced by [JAC2].

#### 3.2.1. Packet Pair based techniques

The packet pair property states that if two packets of the same size are sent back to back then the difference between their arrival times is equal to the propagation delay of a packet over the bottleneck link. More formally define $t_i^j$ to be the arrival time of the jth packet at node i then $t_n^1 - t_n^0 = max_i(S/b_i)$. And therefore $PCB = S/(t_n^1 - t_n^0)$. A detailed proof of the property can be found in [KES]. The reasoning is that the two packets will queue at the bottleneck link and at no other link afterwards and therefore will arrive to the destination with the same spacing as they left the bottleneck link (See Figure 4).

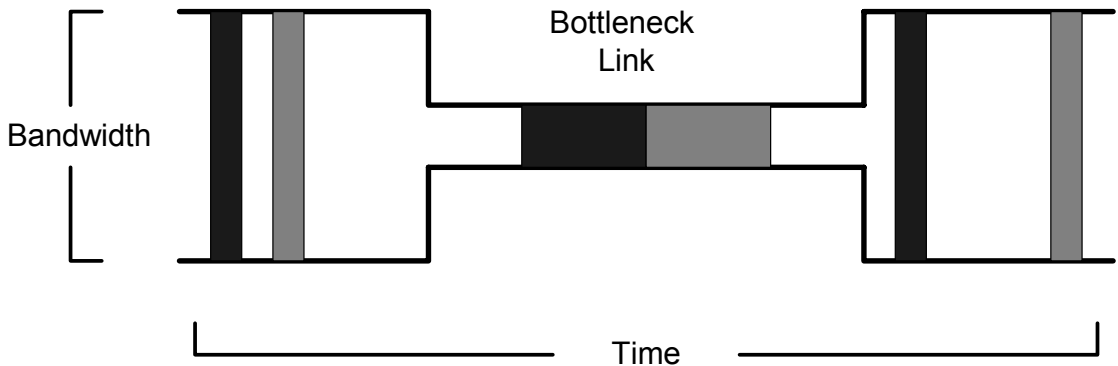

**Figure 4 - Packet Pair Property**

The correctness of the above property depends on several assumptions:
- The routers on the path are store and forward, meaning that a router begins processing a packet only after it's last bit has been received. This property is the common implementation of routers on the internet.

- The two packets are sent sufficiently close to one another. This assumption may be violated only when trying to measure extremely fast links which are not the focus of this paper.
- Both packets take the same route to the destination. It was shown by [PAX] that this assumption is usually true because routes are changed infrequently (the scale is hours).
- There are no multi-channel links. If a link between two nodes has multiple channels then the first and second packets may be sent on different links and this way they may be reordered or their spacing may be changed. Most links are not multi-channel, the exceptions are very fast ATM links and two channel ISDN links.
- There is no cross traffic. Cross traffic may corrupt the results of bandwidth measurement using packet pair. If a cross traffic packet is queued between the two packets of the pair after the bottleneck link then the reported BW will be lower than the actual bottleneck bandwidth. If a cross traffic packet is queued before the first packet in the pair after the bottleneck link then the reported BW will be higher than the actual bottleneck bandwidth. This assumption is usually not true because internet links share traffic from many hosts, therefore various filtering techniques are used to filter out the measurements that were affected by cross traffic.

Below is a summary of related work on the packet pair technique:

The packet pair property was first identified by [JAC1]. [KES] used the packet pair property to measure PAB under fair queuing assumption in the routers along the path. Under fair queuing implementation the queuing delay at the routers is proportional to the size of the packet and therefore as long as the cross traffic is stable, the available bandwidth at the bottleneck link can be measured correctly. Fair queuing is not the common implementation in routers, routers use the simpler FCFS (First come first serve) implementation.

[CAR] presented tools to measure both PCB (bprobe) and PAB (cprobe). bprobe sends pairs of echo[POS] packets to the destination host and measures the difference of arrival times in each pair. The results are then collected and the ones that were affected by cross traffic are filtered out using histograms. The advantage of this technique is that it does not require any cooperation from the receiver host. The disadvantage is that the path must be symmetric, otherwise the reported PCB may belong to a link in the opposite direction. Another problem is the relying on control packets that are not always forwarded in a timely manner by the routers along the path.

[PAX] Solved the problem of links that has multiple channels by sending a bunch of packets instead of only two (Hence the name of the method PBM - packet punch modes). In PBM different bunch sizes where used and the analysis of the gathered results allowed multi-modal distribution so that both multiple channel situations and bottleneck changes over time could be identified.

[DOV] Showed that because of cross traffic, the distribution of packet pair results is multi-modal and that under heavy cross traffic the capacity mode is not the dominant mode. The packet size was also investigated and it was shown that although bigger packets are more immune to noise they also produce more noise because the bigger the packets, the bigger is the chance that cross traffic will be queued between the first and

second packet in the pair. It was also shown that using trains of packets that are long enough produce a single modal distribution of results that converge to a value called ADR (Asymptotic dispersion rate) which is smaller than the bottleneck link capacity. A tool called pathrate was introduced that first found the different distribution modes using packet pair and then found ADR using packet trains, finally it selected the capacity mode from the first stage that was the minimal mode that is still bigger than ADR, this mode was selected as the capacity mode. It should be noted that the last step is a heuristic.
[LAI1] Measures BW passively. Instead of actively sending packet pairs probes are placed on network hosts. The probes identify packets that came from the same source and record their spacing and then send it to a server that groups and analyses the results using density functions. In addition to the regular cross traffic effects that should be filtered out, the situation when the pair packets where not spaced close enough to each other can occur in passive measurement and therefore is identified and filtered out.

### 3.2.2. Delay based techniques

Delay based techniques measure the capacities of all of the links along the path by sending echo packets with limited TTL field.
The filtering of the results of delay based techniques is easier because it is enough to take the minimum of each sample set. The disadvantages are that a large amount of packets has to be sent (Because there is at least one set of measurements for each node along the path), hidden hops may distort the results, and noise may be amplified because of arithmetic operations that must be performed between measurement results that may be inaccurate.

Below is a summary of related work on delay based techniques:
[JAC2] Presents pathchar, a tool that sends a set of packets to a destination node, each with different TTL from 1 to n. This process is repeated with different packet sizes so that finally and a set of graphs is produced each graph shows the delays to a node for different packet sizes. Subtracting a graph of node k from the graph of node k+1 and taking the inverse of the slope of the resulting graph yields the capacity of the link between nodes k and k+1. Filtering out cross traffic affected measurements is done by taking the minimum of each sample set.
[DOW] Shows how the performance of pathchar can be improved in terms of the number of measurements performed. The basic idea is to use adaptive data collection that is to stop probing nodes once their delay value converges.
[JIA]  Presents a variation of the pathchar idea that enables to measure links BW in both directions of the path.
[LAI2] Introduces a new technique called Packet Tailgating. Two steps are performed, the first is similar to pathchar's last step where the sum of bandwidths between the source and destination hops is calculated using linear regression on several packet sizes. The second step performs n measurements in which a very small packet is queued behind a very large packet until the large packet is dropped by one of the routers along the path due to a limited TTL. The second step is performed several times for each TTL value between 1 and n and the minimum value is taken from each sample set. The advantages of this technique are that less packets need to be sent since the second phase does not

require linear regression and therefore there is no need to send packets of different sizes, also ICMP packets are not used (Neither echo nor TTL exceeded), and therefore the technique does not depend on timely processing of such packets in routers.

### 3.3. Bandwidth measurement in CDTP

The Main requirement in CDTP is that the BW measurement must be quick, because it has to be performed each time a new client requests a file. As a result delay based techniques are inadequate because they require to send packets to each node along the path which can consists of more the 15 hops. Packet pair methods are more suitable because packets are sent to the destination host only , the problem with packet pair is that it is very sensitive to noise along the path and to filter out noise a large number of measurements has to be taken.
Our technique uses a packet pair like method, but it's filtering method is the same as in the delay based techniques that is taking a minimum of each sample set is enough.

The technique is made of m identical steps each step is comprised of four stages. The bandwidth calculation is performed on the minimum value of each stage at the end of all steps.
- Stage one of each step sends a packet of size S to the peer host. Upon receiving the packet the peer host sends a reply packet of size S. The sender host calculates the time it took the reply packet to arrive and stores the result in $T_1^i$ variable where i is the index of the step.
- Stage two of each step sends a packet of size 2*S to the peer host. Upon receiving the packet the peer host sends a reply packet of size S. The sender host calculates the time it took the reply packet to arrive and stores the result in $T_2^i$ variable where i is the index of the step.
- Stage three of each step sends a pair of packets back to back of size S each, to the peer host. Upon receiving the second packet the peer host sends a reply packet of size S. The sender host calculates the time it took the reply packet to arrive and stores the result in $T_3^i$ variable where i is the index of the step.
- Stage four of each step sends a pair of packets back to back of size 2*S each, to the peer host. Upon receiving the second packet the peer host sends a reply packet of size S. The sender host calculates the time it took the reply packet to arrive and stores the result in $T_4^i$ variable where i is the index of the step.

Definitions:
- $T_i = Min_{j=1,..,m}(T_i^j)$. – The minimum round trip time of stage i in any of the steps.
- $T_i^{'}$ - The time it took the stage i packets to reach the peer in the step that yields $T_i$.
- $T_i^{\sim}$ - The time it took the stage i packet to travel from the peer back to the source in the step that yields $T_i$.
- $L^*$ - The Bottleneck link.
- $b^*$ - The capacity at the bottleneck link.
- $q^*$ - The queuing delay at the bottleneck link.

Below will be shown how to calculate $b^*$. It should be noted that due to the non deterministic nature of the Internet the values of $q_i$ are different for each measurement and therefore the values of $T_i$ also change with each measurement. In the below calculations it is assumed that the $q_i$ values are constant. The effect of this assumption is minimized by the filtering method that minimizes $T_i^j$ and therefore minimizes the differences between the $q_i$ of the selected measurements.

Using the fact that in all of the stages the reply is the same size and under the above assumption we get:

$$\tilde{T_1} = \tilde{T_2} = \tilde{T_3} = \tilde{T_4} \quad (2)$$

By the packet pair property (proved in Appendix B– Packet Pair Proof) we get:

$$T'_3 - T'_1 = q^* + \frac{S}{b^*} \quad (3)$$

And

$$T'_4 - T'_2 = q^* + 2\frac{S}{b^*} \quad (4)$$

Subtracting (3) from (4) and reordering yields:

$$b^* = \frac{S}{(T'_4 - T'_2) - (T'_3 - T'_1)} \quad (5)$$

Formula (5) gives the bottleneck link capacity but unfortunately none of the $T_i'$ can be measured directly.

From the definitions we get:

$$T'_i = T_i - \tilde{T_i} \quad (6)$$

Now by substituting $T_i'$ with $T_i - \tilde{T_i}$ in (5) and by taking (2) into account we get:

$$b^* = \frac{S}{(T_4 - T_2) - (T_3 - T_1)} = \frac{S}{T_4 + T_1 - T_3 - T_2} \quad (7)$$

Formula (7) gives the bottleneck link bandwidth as a function of $T_i$ which is measured directly.

The queuing delay at the bottleneck can be derived similarly:
By multiplying (3) by 2, subtracting (4) and reordering we get:

$$q^* = 2T'_3 + T'_2 - 2T'_1 - T'_4 \qquad (8)$$

By substituting $T_i'$ with $T_i - \tilde{T_i}$ in (8) and by taking (2) into account we get:

$$q^* = 2T_3 + T_2 - 2T_1 - T_4 \qquad (9)$$

The above method yields the bottleneck link bandwidth by taking the minimum result for each stage and therefore requires much less samples than other packet pair techniques. It also does not require to build up a sample set for each node along the path as required by the delay based techniques.

To avoid measurement errors due to modem compression random bits are sent as payload in the measurement packets.

The above method still cannot measure correctly paths whose bottleneck link has multi-channels.

# 4. CDTP in Detail

## 4.1. Software entities and their roles

CDTP is a session oriented protocol. Each dialog between CDTP client and CDTP server belongs to a certain session with unique ID. Each sub protocol of CDTP defines a set of session types: The Query sub protocol has one session type which is the query session, the Bandwidth sub protocol has also one session type which is the BW session, and the CT protocol has 3 session types, the transfer session the chain session and the file session. The two later session types are internal to the server and their ID is never passed by the protocol.

- Query session is opened between client and server as soon as the client performs successful START_QUERY_SESSION request and remains until the client disconnects either by STOP_QUERY_SESSION command or because of a time out. The number of active Query sessions in the server is exactly the same as the number of active CDTP clients.
- Bandwidth session is opened between server and client or between service client and client when server or service client needs to measure the bandwidth between himself and the client.
- Transfer session is opened between server and client when the client initiates a get request, the transfer session remains until the completion of the get request.
- File session is opened when a is file requested for the first time by any client, the file session is closed after the last client that was requesting the file was serviced.
- Chain Session is opened either by a client that is the first in the file session or by a client that did not find any suitable chain in the file session (suitability of a chain depends on the amount of BW between server and client compared to the amount of BW in the chain, and the amount of data that client at the tail of the chain received compared to the amount of data the requesting client received.

## 4.2. The Protocol

CDTP consists of three sub protocols: Query, Bandwidth and CT.

The Query sub protocol is a subset of FTP commands, including Ls, Pwd, Cd, and Get. All commands except for the Get reply are trivial. Get Reply passes to the client all necessary data to contact the service client which includes: The IP address of the service client, the TCP transfer port at the service client, the transfer session Id of the service client (will be passed to the service client as part of the transfer request for authentication), a transfer session id for the client's current transfer session, the size of the requested file and the BW that the server measured between itself and the client.

The Bandwidth sub protocol is used to measure the bandwidth between the server and client or between the server and a service client, and it is based on the technique described in 3.3.

The CT sub protocol defines the dialog between the client and service client. The client initiates a transfer request from the service client the service client measures the BW between itself and the client and if it is compatible to the BW between the server and the client then the client is serviced.
The protocol commands are:
- Start Download Request – This is the first request that is sent by the client after the TCP session is established. It's parameters are: The Service client's transfer session id (For authentication), the client's transfer session id (will be passed in any subsequent message of the protocol), the position in the file to start transfer from, the BW between the client and the server (to be compared by the service client to the BW between himself and the client), and the number of the client's port for BW measurement replies.
- Transfer Reply – These periodic replies are sent from the service client to the client over the established TCP session. Each reply contains some control parameters and a chunk of the requested file. All reply chunks are of the same size and the last reply chunk is padded with zeros if needed. The parameters are: The transfer session Id, the size of the actual chunk of the file, an indication if this chunk is the last one and the number of the current chunk.
- Transfer Denied reply – Can be sent by the service client over the TCP session at any time during the session it's parameters are: Transfer session id and the denial reason. The possible reasons are: Incompatible link speed or service client transfer session id, service client already serving someone or service client going down.
- Retransmission Request – Sent over UDP by the client to the server if connection to the service client could not be established or was broken. It's parameters are: the client's transfer session id, the number of bytes received by the client so far and an indication if the service client was engaged, this indication is needed so that the server will know if it can direct another client to the service client.
- Retransmission Reply – Sent over UDP by the server to the client to indicate an alternative service client or to indicate a failure, it's parameters are similar to the parameters of the get reply in the query sub protocol.
- Progress Report – Sent over UDP by the client to the server after each successful processing of a transfer reply from the service client, and carries the amount of data that the client received until that point. The server uses the received data when it decides which chain a new client should be assigned to.
- Disengage Command – Sent over UDP from client to the server to indicate that the transfer session has ended and no clients will be served over the session (if the client was the last in the chain then the chain is closed). The passed parameters are the amount of data received by the client and an indication if the service client was engaged by the client.

## *4.3. Scenarios*

### 4.3.1. Acronyms in the Diagrams

- A solid line means a message between different hosts.
- A dotted line means a command inside a host.
- S Query – Component responsible for the query sub protocol in the server.
- S Control - Component responsible for the control messages of the CT sub protocol in the server.
- S Transfer - Component responsible for the data transfer messages of the CT sub protocol in the server.
- S BW – Component responsible for the BW sub protocol in the server.
- C Client – Client that requests data..
- C Service – Client that serves another client.

### 4.3.2. First Client Flow

The figure below shows a flow for a client that requests a file, and a new chain has to be created for that client:

1. The client issues a get request to the server.
2. The Server measures the BW between himself and the client and creates a chain with the client as the head of the chain.
3. The Server replies the Get command where it specifies it's own address as the address to which the client should connect to receive the file it requested.
4. Upon receiving the reply, the client starts requesting the file from the server (the file is requested in chunks so that other client's could be served by the first client).
   During data retrieval the client sends progress reports to the server.
5. At the end the client issues a disengage command to the server and if no clients were added to the chain, then the server removes the chain.

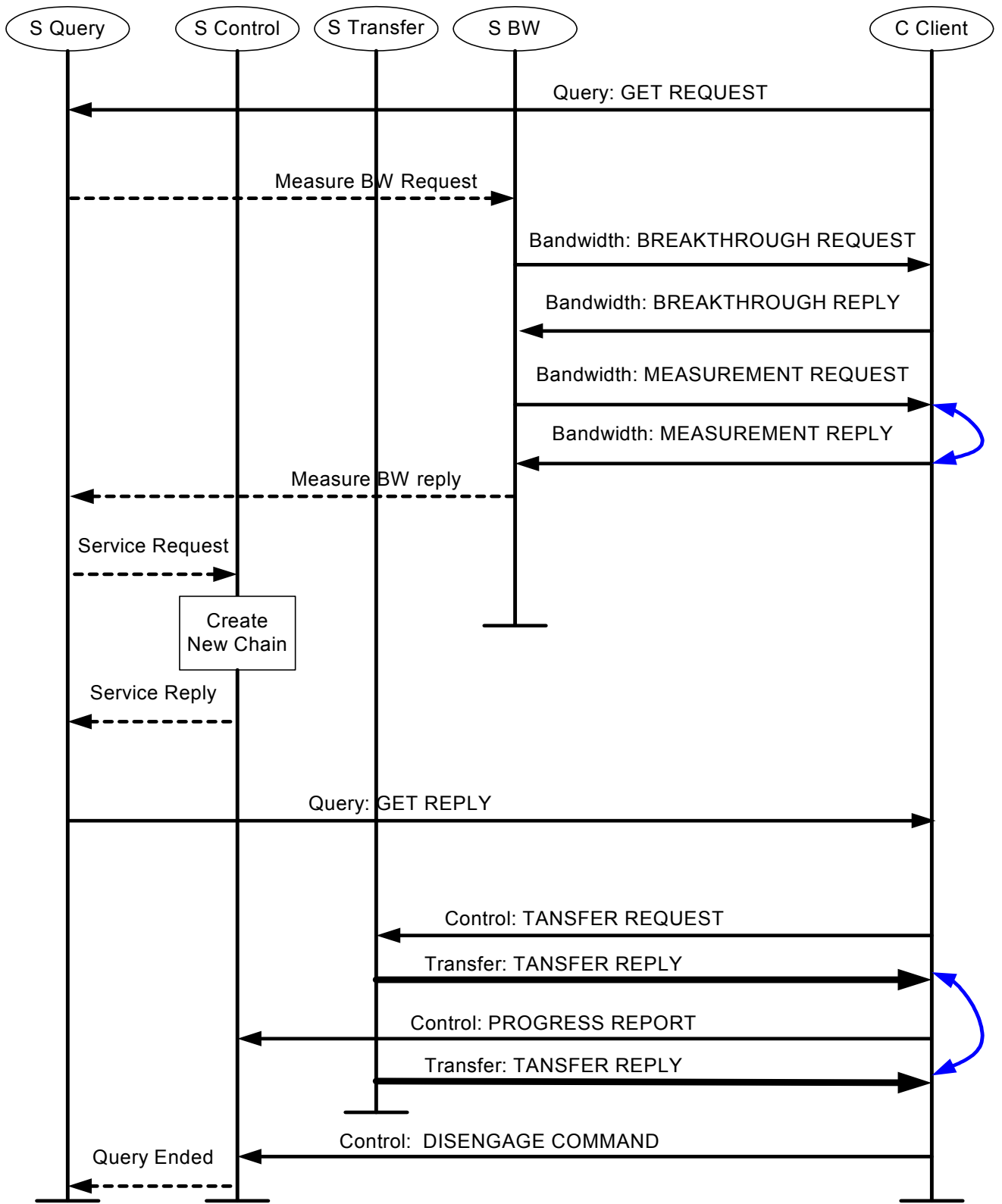

**Figure 5 – First Client Flow**

### 4.3.3. General Client flow

The figure below shows a flow for a client that requests a file that already being transferred to other clients:

1. The client issues a get request to the server.
2. The Server measures the BW between himself and the client, finds a chain with compatible link speed and places the client on the tail of the found chain.
3. The Server replies the Get command where it specifies the address of the last client in the found chain as the address to which the client should connect to receive the file it requested.
4. Upon receiving the reply from the server the client requests the service client (whose address and port were given to the client in the get reply), to pass him the file.
5. The service client measures the BW between himself and the requesting client, sees that it compatible to his own and then starts sending data to the client.
6. The rest is like in the previous case.

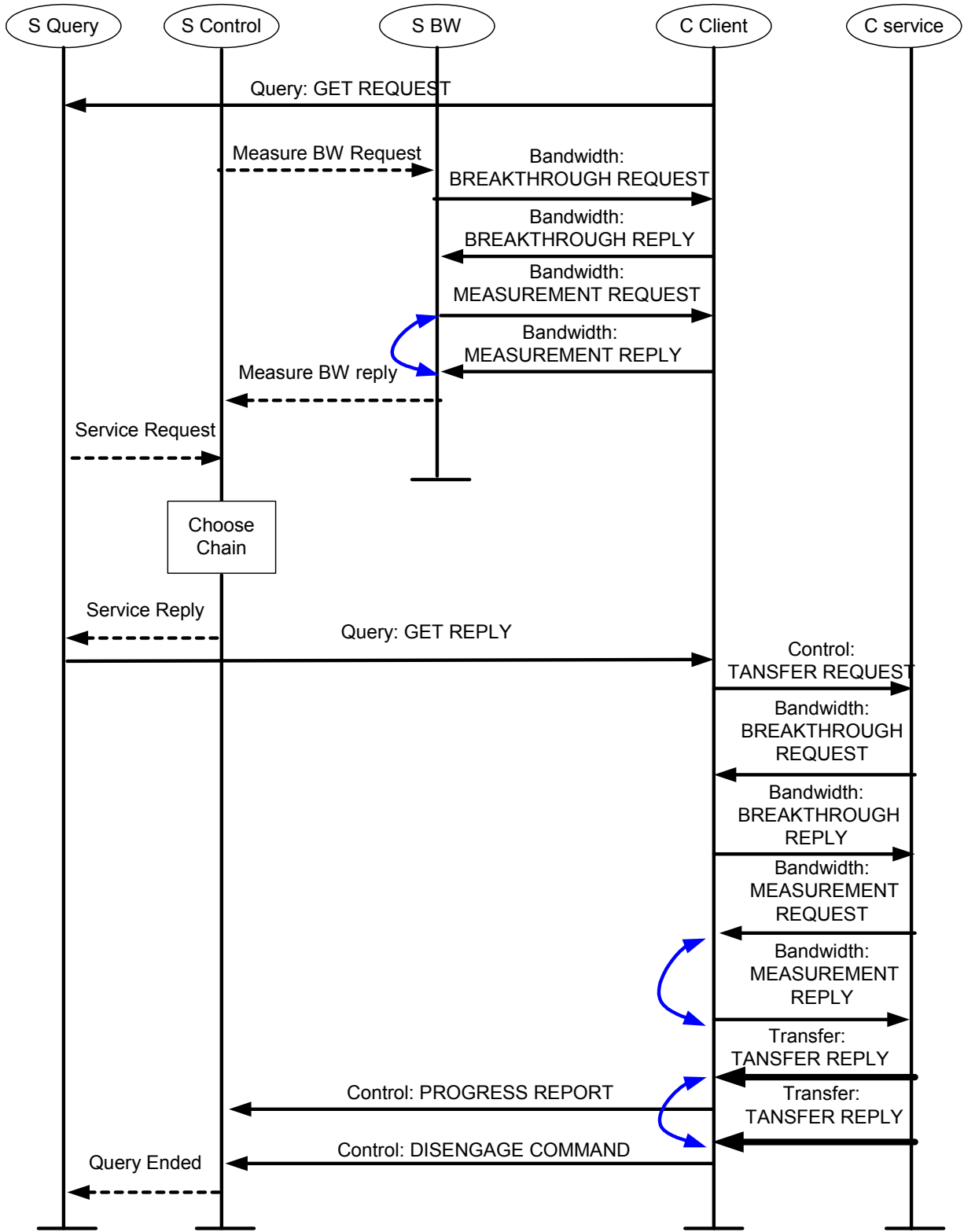

**Figure 6 – General Client Flow**

### 4.3.4. Reconnecting a Broken Chain

The figure below shows a retransmission flow:

1 – A transfer to a client was broken either because of an explicit message from the service client or because of a timeout.
2 – The client issues a retransmission request to the server.
3 – The server finds a suitable chain for the client (The chain is chosen according to two parameters: 1 – The BW to the client 2- The amount of data the client already received).
4 – A redirection reply is sent to the client,.
5 – The rest is similar to the last two scenarios.

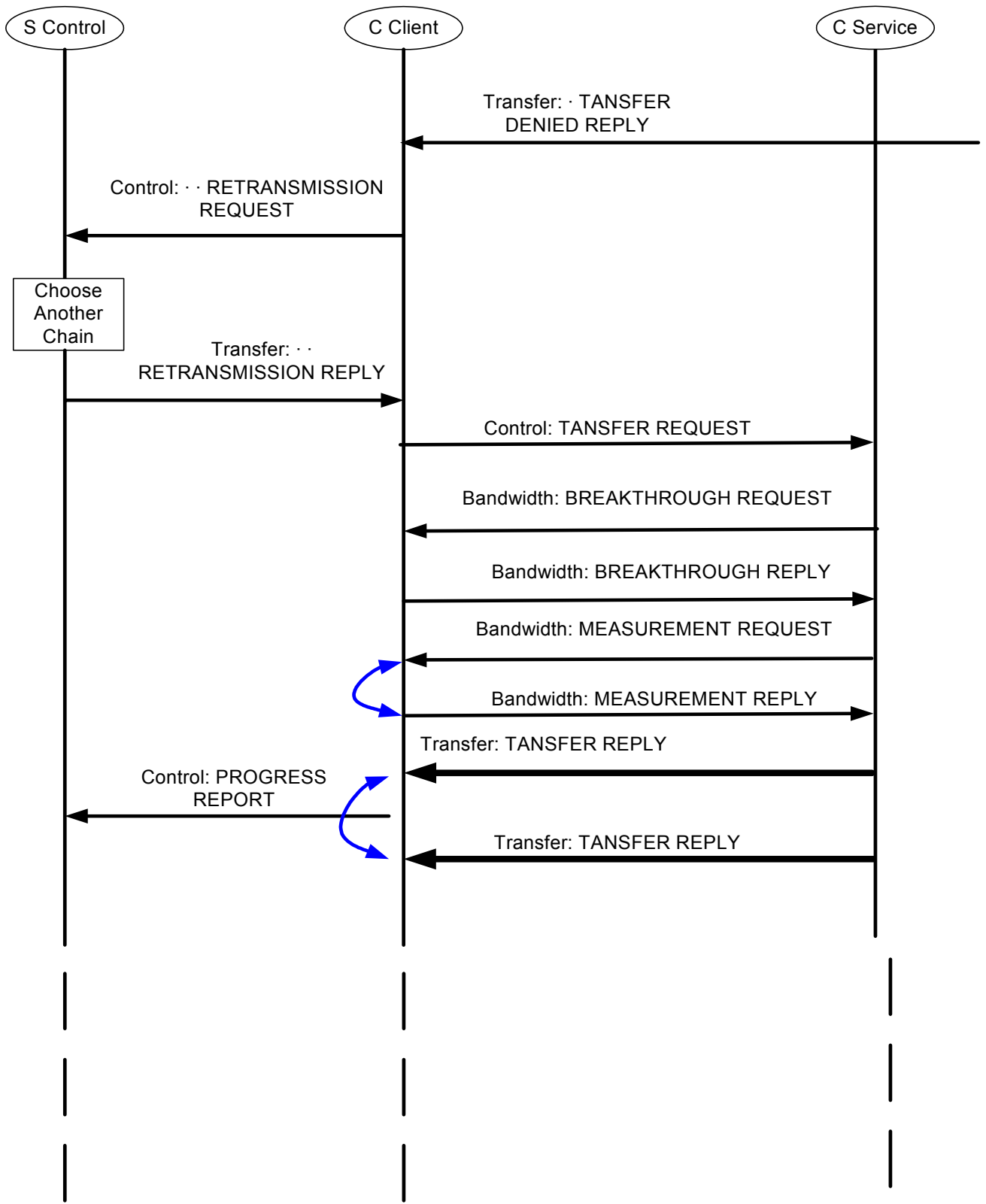

**Figure 7 - Reconnecting a Broken Chain**

# 5. Results of experiments

## 5.1. CDTP experiments

The experiment was conducted in a lab with one computer that acted as a server and 3 other computers that acted as clients. The clients resided on the same 100MB Ethernet and therefore the access speed between them was high. The server was behind a wireless modem that was configured with 120Kbps, 240Kbps and 480Kbps speeds. In each of the tests a file was requested by all of the clients almost simultaneously (There was a 5 seconds difference between the requests. The requests were always been in the same order: Client 1 , client 2 and Client 3.) using CDTP and using FTP.

The table below summarizes the results of the experiments:

| Client # | Server Speed (Kbps) | File Size (Bytes) | CDTP Transfer Time (Sec) | FTP Transfer Time (Sec) | CDTP Rate (bps) | FTP Rate (bps) |
|---|---|---|---|---|---|---|
| 1 | 120 | 561282 | 55 | 101 | 81641 | 44457 |
| 2 | 120 | 561282 | 48 | 108 | 93547 | 41576 |
| 3 | 120 | 561282 | 43 | 106 | 104424 | 42360 |
| 1 | 240 | 561282 | 28 | 50 | 160366 | 89805 |
| 2 | 240 | 561282 | 23 | 51 | 195228 | 88044 |
| 3 | 240 | 561282 | 19 | 49 | 236329 | 91637 |
| 1 | 480 | 4567025 | 88 | 233 | 415184 | 156807 |
| 2 | 480 | 4567025 | 81 | 235 | 451064 | 155473 |
| 3 | 480 | 4567025 | 72 | 232 | 507447 | 157483 |

**Table 1 – CDTP vs FTP**

In all of the CDTP experiments client 2 downloaded the file faster than client 1 and client 3 was faster than client 2. The reason is that client 2 started the downloaded 5 seconds after client 1 and client 3 started the download 5 seconds after client 2 and therefore part of the data that clients 2 and 3 were using was already cashed at clients 1 and 2 respectively.

In all of the FTP experiments client 2 was the slowest. The reason is that in FTP all 3 clients got the file directly from the server, and client 2 download was overlapped by either client 1 or client 3 or both during it's entire download period.

The bandwidth utilisation in CDTP was not optimal because of the bandwidth measurement phase at the beginning of the query. The utilisation was better in the last experiment (68% in the first, 66% in the second and 86% in the third) because the transferred file was larger and therefore the effect of the bandwidth measurement which is constant in size was smaller.

### 5.2. Bandwidth measurement experiments

The bandwidth measurement was tested both in lab conditions using a wireless modem and in internet conditions using a standard modem.

In the lab a path of 2 hops was measured: The first hop was a 100 Mbps fast Ethernet, the second hop was the bottleneck hop and it was tested with bandwidth of 60Kbps, 120 Kbps, 240 Kbps and 480 Kbps. The base packet size was 600 bytes and the number of steps per each measurement was 3.

The table below summarizes the results of the lab experiments:

| Link Speed(bps) | Measurement # | Calculated speed (bps) | Accuracy (%) |
| --- | --- | --- | --- |
| 60000 | 1 | 60052 | 99.91 |
| 60000 | 2 | 63879 | 93.92 |
| 60000 | 3 | 60039 | 99.93 |
| 60000 | 4 | 59980 | 99.96 |
| 60000 | 5 | 59998 | 99.99 |
| 120000 | 1 | 120187 | 99.84 |
| 120000 | 2 | 120769 | 99.36 |
| 120000 | 3 | 157750 | 76.0 |
| 120000 | 4 | 120131 | 99.89 |
| 120000 | 5 | 119968 | 99.97 |
| 240000 | 1 | 238895 | 99.53 |
| 240000 | 2 | 240641 | 99.73 |
| 240000 | 3 | 240469 | 99.80 |
| 240000 | 4 | 240006 | 99.99 |
| 240000 | 5 | 241595 | 99.33 |
| 480000 | 1 | 473171 | 98.57 |
| 480000 | 2 | 474465 | 98.84 |
| 480000 | 3 | 478508 | 99.68 |
| 480000 | 4 | 478189 | 99.62 |
| 480000 | 5 | 483804 | 99.21 |

**Table 2 – Lab bandwidth measurements**

As can be seen in table 2 the measurement was very accurate except for one measurement with modem speed 120 kbps. The reason is probably a burst of cross traffic that disrupted the measurement. Taking a larger number of steps should have fixed the inaccuracy.

In the internet experiments the following setups were used:
Setup 1 – Two PCs were both connected with a 38.4Kbps modem link to each other over a path of 12 hops.
Setup 2 – Two PCs were connected over 16 hops. The first PC that was the sender was connected using a 38.4Kbps modem and the second that was the receiver was connected using a 384Kbps fractional E1 Link.
Setup 3 – The same as setup two but with the roles of the sender and receiver reversed. In the reverse direction the number of hops was 15.

The number of hops was calculated using the trace-route utility in the PC.
The actual link speed of the modem connections was not exactly known so it was approximated by correlating 3 sources: 1 –Pings to the nearest router 2 – An internet site that measures BW by downloading large amounts of data to the browser [BWP] 3 – Our own measurements. In all cases the measured BW (Both uplink and downlink) was around 26 kbps so it was used as the reference speed for the measurements.
Because of the internet instability the number of steps in each measurement was increased to 10.

| Setup # | Measurement # | Calculated speed (bps) | Accuracy (%) | Packet pair 1 speed (bps) | Packet pair 2 speed (bps) |
|---|---|---|---|---|---|
| 1 | 1 | 25420 | 97.76 | 23552 | 24451 |
| 1 | 2 | 22473 | 86.43 | 24883 | 23617 |
| 1 | 3 | 26320 | 98.78 | 23183 | 24652 |
| 1 | 4 | 25427 | 97.79 | 23061 | 24186 |
| 1 | 5 | 22789 | 87.65 | 24976 | 23833 |
| 2 | 1 | 27296 | 95.25 | 25171 | 26191 |
| 2 | 2 | 26036 | 99.86 | 25912 | 25974 |
| 2 | 3 | 26326 | 98.76 | 25949 | 26136 |
| 2 | 4 | 26503 | 98.10 | 25754 | 26123 |
| 3 | 1 | 26621 | 97.66 | 25423 | 26008 |
| 3 | 2 | 26443 | 98.32 | 25190 | 25801 |
| 3 | 3 | 26600 | 97.74 | 25210 | 25886 |
| 3 | 4 | 26719 | 97.30 | 25095 | 25882 |
| 3 | 5 | 26403 | 98.47 | 25558 | 25974 |
| 3 | 6 | 26601 | 97.74 | 24977 | 25763 |

**Table 3 – Internet bandwidth measurements**

Considering the internet instabilities the measurement results were fairly accurate, with the exception of measurements 2 and 5 in the first setup. However these inaccuracies can be detected:

From formulas (3) and (4) we get:

$$0 < ((T_4 - T_2) - (T_3 - T_1)) < \frac{1}{2}(T_4 - T_2) < T_3 - T_1 < T_4 - T_2 \quad (10)$$

Therefore:

$$\frac{S}{(T_4 - T_2) - (T_3 - T_1)} > \frac{2S}{T_4 - T_2} \quad (11)$$

And:

$$\frac{2S}{T_4 - T_2} > \frac{S}{T_3 - T_1} \quad (12)$$

Thus we get that the value in the third column in table 3 should be bigger than the value of the fifth column which in turn should be bigger than the value of the sixth column. The intuitive explanation of formula (11) is that the calculated bandwidth value should be greater when the delay is negated and the explanation for formula (12) is that when the packet is bigger than the delay inside the router has a lesser effect on the bandwidth calculation.

Now returning to measurements 2 and 5 in setup 1 it can be seen that formulas (7) and (8) do not hold and therefore the BW calculation is erroneous. A possible adaptive way to handle such situation is to continue measuring until the inequalities hold.

## 6. Summary and Future Research

In this work we presented CDTP, a distributed transfer protocol that uses the unused portion of the internet clients bandwidth to relax the computation and bandwidth requirements of internet servers and unlike Napster and Gnutella CDTP does not require that clients store the downloaded files for periods longer then necessary to receive them.
A bandwidth measurement technique was presented that combined the ideas of the packet pair techniques with the ideas of delay based techniques to achieve accurate bandwidth measurements while using small amount of measurement packets.
A way to extend CDTP is to use its ideas to transfer other kinds of traffic like HTTP or streaming multimedia. The main obstacle with HTTP is that it's traffic consists of much smaller data units and therefore the bandwidth measurement methods should be more efficient.
Another improvement could be to add the ability to detect multi-path bottleneck links and still send a small amount of bandwidth measurement packets.

# 7. Appendix A – The Protocol

## 7.1. Query Protocol

### 7.1.1. General

The query protocol is used between the client and server for high-level control.
Underlying Protocol: UDP.
Ports: Server uses Port 6000. Clients use transient ports.
Protocol Id: 10.

### 7.1.2. Layout

- For each Packet:

Byte0– Protocol Id.
Byte 1 – Operation Id.
Bytes 2 – 3: Number of fields.
Byte 4: Payload Indication.
Rest of bytes: The fields.

- For each Field:

Bytes 0 – 3: Length of data.
Rest of bytes: Data.

Layout:

| 0 | 1 | 2 – 3 | 4 | 5 – 8 | ------- Rest Of Bytes ------ | | |
|---|---|---|---|---|---|---|---|
| Protocol Id(0) | Op Id | Fields Number | Payload ? | Field 1 Size | Field 1 Data | ……. | Field N Data |

### 7.1.3. Operations

- START QUERY SESSION REQUEST
- Initiator: Client
- Op Id: 0.
- Field 1 – Client UDP Query Port. Field1 Size: 2 bytes.
- Field 2 – Service Client TCP Transfer Port. Field2 Size: 2 bytes.
- Field 3 – Client UDP Control Port. Field3 Size: 2 bytes.
- Field 4 – Client UDP BW Peer 2 Port. Field5 Size: 2 bytes.
- Description: A request to start a query session, the passed port numbers:
  - Client UDP Query Port  - Will be used by the server to communicate to the client throughout the session of the Query protocol.
  - Service Client TCP Transfer Port: With this port the client acts as service client and serves other clients.
  - Client UDP Control Port – Will be used by the server to communicate to the client throughout a session of Control Protocol.

- Port of the client's Peer 2 BW Protocol.

- START QUERY SESSION REPLY
- Initiator: Server
- Op Id: 1.
- Field 1 – Session Handle. Field Size: 4 bytes.
- Field 2 – Success Indication. Field Size: 1 byte.
- Description: A reply to a start query session request by a client. If request is granted then a session Id is returned.

- STOP QUERY SESSION COMMAND
- Initiator: Server or Client.
- Op Id: 2.
- Field 1 – Session Handle. Field Size: 4 bytes.
- Description: An indication that the query session is closed by one of the sides.

- Operation: LS REQUEST
- Initiator: Client.
- Op Id – 3.
- Field 1 – Session Handle. Field Size: 4 bytes.
- Description: Similar to the UNIX ls command.

- Operation: LS REPLY
- Initiator: Server.
- Op Id – 4.
- Field 1 – Session Handle. Field Size: 4 bytes.
- Field 2 – Success Indication. Field Size: 1 byte.
- Each following field – Name of file or directory. Field Size: Variable.
- Description: The list of files and directory names from the current directory associated to the session.

- Operation: PWD REQUEST
- Initiator: Client.
- Op Id – 5.
- Field 1 – Session Handle. Field Size: 4 bytes.
- Description: Similar to the pwd UNIX command.

- Operation: PWD REPLY
- Initiator: Server.
- Op Id – 6.

- Field 1 – Session Handle. Field Size: 4 bytes.
- Field 2 – Success Indication. Field Size: 1 byte.
- Field 3 – Current path. Field Size: Variable.
- Description: The path of the current directory related to this session.

- Operation: CD REQUEST
- Initiator: Client.
- Op Id – 7.
- Field 1 – Session Handle. Field Size: 4 bytes.
- Field 2 – Directory Name. Field Size: Variable.
- Description: Similar to the UNIX cd command.

- Operation CD REPLY
- Initiator: Server.
- Op Id – 8.
- Field 1 – Session Handle. Field Size: 4 bytes.
- Field 2 – Success Indication. Field Size: 1 byte.
- Description: Replies if the CD command succeeded.

- Operation: GET REQUEST
- Initiator: Client.
- Op Id – 9.
- Field 1 - Session Handle. Field Size: 4 bytes.
- Field 2 – File Name. Field Size: Variable.
- Description: A request to get a file from the server.

- Operation: GET REPLY
- Initiator: Server
- Op Id – 10.
- Field 1 – Session Handle. Field Size: 4 bytes.
- Field 2 – Success Indication. Field Size: 1 byte.
- Field 3 – IP of the service client. Field size: 4 bytes.
- Field 4 – Port number of the TCP transfer port of the service client. Field Size 2 bytes.
- Field 5 – Service Client's Transfer Session Id. Field size: 4 bytes.
- Field 6 – Transfer session Id: 4 bytes.
- Field 7 – Size of file. Field Size: 4 bytes.
- Field 8 – Link Speed. Field Size 4 Bytes.

- Description: A reply to a get request. Provides the address of the actual server to get the data from and a unique identifier for this transfer session and the size of the requested file. The Link Speed between the client and server is provided also.

## 7.2. Control & Transfer Protocol

### 7.2.1. General

The control & transfer protocol is used between the client and server and between clients for controlling and transferring the requested data.
Underlying Protocol: UDP, TCP
Ports: UDP Server uses Port 6000, clients use transient ports, Service Clients use transient ports.
Protocol Id: 20.

### 7.2.2. Layout

- For each Packet:

Byte 0 – Protocol Id.
Byte 1 – Operation Id.
Bytes 2 – 3: Number of fields.
Byte 4: Payload Indication.
Rest of bytes: The fields.

- For each Field:

Bytes 0 –3: Length of data.
Rest of bytes: Data.

Layout:

| 0 | 1 | 2 – 3 | 4 | 5 – 8 | ------- Rest Of Bytes ------ | | |
|---|---|---|---|---|---|---|---|
| Protocol Id(0) | Op Id | Fields Number | Payload ? | Field 1 Size | Field 1 Data | ……. | Field N Data |

### 7.2.3. Operations

- START TANSFER REQUEST
- Protocol: TCP.
- Initiator: Client
- Recipient: Server or a Service Client
- Op Id: 0.
- Field 1 – Service Client's Transfer Session Id. Field Size: 4 bytes.
- Field 2 – Transfer Session Id. Field Size: 4 bytes.
- Field 3 – Start Position. Field Size: 4 bytes.
- Field 4 – Chunk Size. Field Size: 4 Bytes.

- Field 5 – Requested Link Speed: 4 bytes.
- Field 6 – BW Protocol Peer 2 UDP Port: 2 bytes.
- Description: A request to start transferring the data.
    - The start position indicates from what location in the file to start the transfer (It is needed when clients reassigned to alternate server after completing part of the transfer).
    - Chunk size indicates the amount of data sent with each transfer. The requested link speed is also provided, so that the Service Client will measure the link speed between itself and the client and if it is not within acceptable range, the request will be denied.

- TRANSFER REPLY
- Protocol: TCP.
- Initiator: Service Client or the Server
- Recipient: Client
- Op Id: 1.
- Field 1 – Transfer Session Id. Field Size: 4 bytes.
- Field 2 – Actual chunk size. Field Size: 4 bytes.
- Field 3 – Is last. Field Size: 1 byte.
- Field 4 – Chunk number: Field Size 4 bytes.
- Field 5 – Data Chunk: Field Size Same as "Chunk size" field indicated in START TANSFER REQUEST.
- Description: A transfer of requested data.
    - All the chunks are of the size indicated in the request packet, When less data than Chunk size is passed (For example the last chunk or if the server does not have all the data yet ) the remainder is padded with zeros, the size of actual data is indicated in the actual data field.
    - The is last field indicates if this transfer is the last.

- TANSFER DENIED REPLY
- Protocol: TCP.
- Initiator: Service client or Server
- Recipient: Client
- Op Id: 2.
- Field 1 – Transfer Session Id. Field Size: 4 bytes.
- Field 2 – Reason. Field Size: 1 bytes.

Description: Denial to service a request, the reason values are 1: Incompatible link speed. 2: Transfer Session Id of the service client was incompatible. 3: Service client already serving. 4: Serving Client going down. 5: Chunk size too big  6: Internal

- RETRANSMISSION REQUEST
- Protocol: UDP.
- Initiator: Client

- Recipient: Server
- Op Id: 3.
- Field 1 – Transfer Session Id. Field Size: 4 bytes.
- Field 2 – Size Received. Field Size: 4 bytes.
- Field 3 – Service client occupied. Field Size: 1 Byte.

Description: The link between the client and the Service Client, is broken, or cannot be established therefore the client requests the server to give it an alternate source. Size received indicates the amount of data received by the client so far.
The Service Client occupied field indicates if the service client has accepted the connection attempt from the client or not (It is important for the server to know this because it should not assign new clients to the service client's chain if it accepted the connection).

- RETRANSMISSION REPLY
- Protocol: UDP.
- Initiator: Server
- Recipient: Client
- Op Id: 4.
- Field 1 – Transfer Session Id. Field Size: 4 bytes.
- Field 2 – Success Indication. Field Size: 1 bytes.
- Field 3 – IP of alternate service client. Field Size: 4 bytes.
- Field 4 – Port number of the TCP transfer port of the service client. Field Size 2 bytes.
- Field 5 – New Service Client Transfer Session Id. Field Size: 4 bytes.

Description: A reply from the server that indicates the alternate server to that will supply the rest of the data for the client, a new chain is used because the alternate server will be located in the new chain (The old one is broken) and therefore a new Service Client Transfer Session Id is given. The transfer session Id remains the same.

- PROGRESS REPORT
- Protocol: UDP.
- Initiator: Client
- Recipient: Server
- Op Id: 5.
- Field 1 – Transfer Session Id. Field Size: 4 bytes.
- Field 2 – No of  bytes received. Field Size: 4 bytes.

Description:  A report indicating the number of bytes received by the client so far. This info will be used by the server when it decides where to redirect clients that need retransmission, and where to assign new clients.

- DISENGAGE COMMAND
- Protocol: UDP.
- Initiator: Client
- Recipient: Server

- Op Id: 6.
- Field 1 – Transfer Session Id. Field Size: 4 bytes.
- Field 2 – No of bytes received. Field Size: 4 bytes.
- Field 3 – Service client occupied. Field Size: 1 Byte.

Description: An indication by the client that it is disengaging and therefore it's chain is closed. The No of bytes received is needed for statistics.

The Service Client occupied field indicates if the service client has accepted the connection attempt from the client or not (It is important for the server to know this because it should not assign new clients to the service client's chain if it accepted the connection).

### 7.3. Bandwidth Protocol

The bandwidth protocol used by two peers to determine the speed of the links between them.

The protocol works in the following way:

1 – Peer 1 sends "breakthrough" packet to peer 2. The reason of this packet is to update caches in the routers between the two peers.
2 – Peer 2 replies that it received the "breakthrough" packet.
3 – Peer 1 sends a packet to peer 2, before sending the packet peer 1 measures the current time and put it in the sent packet.
4 – Peer 2 sends the same packet exactly to Peer 1.
5 – When Peer 1 receives the reply it measures the current time, subtracts from it the time of sending (This time is written in the received packet) and stores the result as T1.
6 – Peer 1 sends a packet which is # times bigger than the previous one (# is the division factor), with an indication to peer 2 to divide the size of the returned packet by #.
7 – Peer 2 receives the packet, divides it's size by # and sends it back to peer1.
8 – When Peer 1 receives the reply packet it calculates the time the second packet travelled back and forth and stores it in T2.
9 – Peer 1 sends a pair of packets back to back, the packet sizes are the same as the size of the packet in stage 3. Before sending the first packet of the pair peer 1 measures the current time and puts it in the sent packets.
10 – Upon receiving the second packet of the pair peer 2 sends a single packet of the same size back to peer 1.
11 - When Peer 1 receives the reply it calculates the time the packets travelled back and forth and stores it in T3.
12 – Peer 1 sends a pair of packets back to back, the packet sizes are the # times bigger than the original packet. Before sending the first packet of the pair peer 1 measures the current time and puts it in the sent packets.
13 – Upon receiving the second packet of the pair peer 2 divides the packet's size by # and sends a single packet back to peer 1.
14 - When Peer 1 receives the reply it calculates the time the packets travelled back and forth and stores it in T4.
- The above procedure is performed several times and then the minimum of each $T_i$ is taken and the bottleneck bandwidth is calculated using formula (7).

- Note that this calculation is correct even when the links is asymmetrical, or when the packets travel back in a different route.

Underlying Protocol: UDP.
Ports: Peer 1 uses transient ports, Peer 2 use transient port.
Protocol Id: 30.

### 7.3.1. Layout

- BREAKTHROUGH REQUEST:
- Initiator: Peer 1.
- Byte 0 – Protocol Id: Field Size 1 Byte.
- Byte 1 – Packet type – 0 ( Breakthrough request ).
- Byte 2 – 5 Identifying token: Field size: 4 Bytes.
- Byte 6 – 7 Peer Port: Field size: 2 Bytes.

- BREAKTHROUGH REPLY:
- Initiator: Peer 2.
- Byte 0 – Protocol Id: Field Size 1 Byte.
- Byte 1 – Packet type – 1 ( Breakthrough reply ).
- Byte 2 – 5 Identifying token: Field size: 4 Bytes.
- Byte 6 – 7 Peer Port: Field size: 2 Bytes.

- MESUREMENT REQUEST:
- Initiator: Peer 1.
- Byte 0 – Protocol Id: Field Size 1 Byte.
- Byte 1 – Packet type – 2 ( Measurement request ) , 4 (First Packet in a Packet Pair), 5 (Second Packet in a packet pair). Field size: 1 Byte.
- Byte 2 – 5 Identifying token. Field size: 4 Bytes.
- Byte 6 – 7 – Peer Port. Field size: 2 Bytes.
- Bytes 8 – 11 Sequence number. Field size: 4 Bytes.
- Bytes 12 – 19 – Send Time. Field size: 8 Bytes.
- Byte 20 – Multiplication factor. Field size: 1 Byte.
- Byte 21 – 24 Base Payload size. Field size: 4 Bytes.
- Rest of bytes are just payload that makes the packet # times bigger than the base payload size, where # is the value of the multiplication factor field.

- MESUREMENT RELPY:
- Initiator: Peer 2.
- Byte 0 – Protocol Id: Field Size 1 Byte.
- Byte 1 – Packet Type – 3 ( Measurement reply ).
- Byte 2 – 24 – Exactly the same as in measurement request (Except for the port field).

- Rest of bytes are just payload that makes the whole packet be of the size indicated in the previous field.

### *7.4. Batch Protocol*

#### 7.4.1. General
The Batch protocol is not a part of CDTP.
The Batch protocol is used to control client operations remotely.

Underlying Protocol: UDP
Ports: All CDTP clients use Port 7000, the controlling application uses port 7001.
Protocol Id: 40.

#### 7.4.2. Layout
- For each Packet:
Byte 0 – Protocol Id.
Byte 1 – Operation Id.
Byte 2 – Data.

#### 7.4.3. Operations
- BEGIN REQUEST
- Protocol: UDP.
- Initiator: Controlling application
- Recipient: Client
- Op Id: 0.
- Data field: Always 0.
- Description: A request from the controlling application to start a Get operation.

- BEGIN REPLY
- Protocol: UDP.
- Initiator: Client
- Recipient: Controlling application
- Op Id: 1.
- Data field: 0 – If the client failed to begin the operation, 1 otherwise. .
- Description: A reply from the client to the controlling application that indicates if the Get operation began.

- BEGIN RESULT
- Protocol: UDP.
- Initiator: Client
- Recipient: Controlling application
- Op Id: 2.
- Data field: 0 – If the get operation failed, 1 otherwise. .

- Description: A reply from the client to the controlling application that indicates if the Get operation succeeded.

## 8. Appendix B– Packet Pair Proof

In this section the packet pair property is proved under the assumptions and definitions made in 3. The packet pair property was first proved by [KES] under fair queuing assumption in routers.

Definitions:
- $P_1,...,P_{n+1}$ – The nodes along the measurement path.
- $L_1,...,L_n$ – The links that connect the nodes along the measurement path.
- $l_1,...,l_n$ – The latencies over the links along the measurement path.
- $c_1,...,c_n$ – The link capacities along the measurement path.
- $q_1,...,q_{n+1}$ – The queuing delays at the nodes along the measurement path due to cross traffic (Note that $q_{n+1} = 0$).
- S – The size of the sent packets.
- $c^*$ and $q^*$ - A pair $c_k$ and $q_k$ that fulfils: $q_k + S / c_k = \text{Max}_{I = 1,..,n} \{ q_i + S / c_i \}$.

*Theorem*
Suppose a pair of packets F(irst) and (secon)D of size S are sent back to back from $P_1$ to $P_{n+1}$.
Define $T_f$ as the arrival time of F to $p_{n+1}$ and $T_d$ as the arrival time of D to $p_{n+1}$.
Then:

$$T_d - T_f = q^* + \frac{S}{c^*} \qquad (13)$$

*Proof:*
The proof is done by induction on n.

The base of the induction n = 1:
Denote the time when the last bit of F was sent over $L_1$ as T. Then:

$$T_f = T + l_1 \qquad (14)$$

Since we assume that all nodes are store and forward it means that $P_1$ will not treat D until the last bit of F is sent to $P_2$. And therefore:

$$T_d = T + q_1 + \frac{S}{c_1} + l_1 \qquad (15)$$

From (14), (15) and the fact that the single link is the bottleneck link we get (13).

The induction step from n-1 to n:
Define $T_f^{'}$ as the arrival time of F to $p_n$ and $T_d^{'}$ as the arrival time of D to $p_n$.
Define $c^{'}$ and $q^{'}$ as the capacity and queuing delay at the bottleneck on the path between $P_1$ and $P_n$.
Then by the induction hypothesis:

$$T_d^{'} - T_f^{'} = q^{'} + \frac{S}{c^{'}} \qquad (16)$$

Two cases are possible: $L_n$ is the bottleneck and then $q^{*} = q_n$ and $c^{*} = c_n$ or $L_n$ is not the bottleneck and then $q^{*} = q^{'}$ and $c^{*} = c^{'}$.
In the first case, since the last link is the bottleneck link we get:

$$q^{'} + \frac{S}{c^{'}} < q_n + \frac{S}{c_n} \qquad (17)$$

Combining (16) and (17) gives:

$$T_d^{'} < T_f^{'} + q_n + \frac{S}{c_n} \qquad (18)$$

Meaning that D arrives at $P_n$ before $P_n$ has finished sending F over $L_n$.
Using similar reasoning as in the induction's base we get that:

$$T_d - T_f = q_n + \frac{S}{c_n} = q^{*} + \frac{S}{c^{*}} \qquad (19)$$

In the second case, since the last link is not bottleneck link we get:

$$q^{'} + \frac{S}{c^{'}} > q_n + \frac{S}{c_n} \qquad (20)$$

Combining (16) and (20) gives:

$$T'_d > T'_f + q_n + \frac{S}{c_n} \qquad (21)$$

Meaning that D arrives at $P_n$ after $P_n$ has finished sending F over $L_n$ and therefore:

$$T_d = T'_d + q_n + \frac{S}{c_n} \qquad (22)$$

The above formula is always true for F and therefore:

$$T_d - T_f = (T'_d + q_n + \frac{S}{c_n}) - (T'_f + q_n + \frac{S}{c_n}) = T'_d - T'_f = q^* + \frac{S}{c^*} \qquad (23)$$

In the proof it was assumed that there is a single bottleneck link along the path. The assumption was made to ease the exposition. The proof is easily extended to the case of multiple bottleneck links.

# References

[DSL]  www.dslforum.org

[FTP] File Transfer Protocol. J. Postel, J.K. Reynolds. Oct-01-1985. RFC - 0959.

[AKA] www.akamai.com

[NAP] http://opennap.sourceforge.net/napster.txt

[GNU] http://www.stanford.edu/class/cs244b/gnutella_protocol_0.4.pdf - A distributed systems course home page at Stanford university by Prof David Cheriton.

[COM] Internetworking with TCP/IP By Douglas E. Comer, Prentice Hall 1995

[Jac1] Van Jacobson. Congestion avoidance and control. In SIGCOMM '88 Conference Proceedings, pages 314-329, Stanford, CA, USA, August, 1998.

[Jac2] Van Jacobson. Pathchar - a tool to infer characteristics of Internet paths. Presented at the Mathematical Sciences Research Institute (MSRI); Slides available from ftp://ftp.ee.lbl.gov/pathchar/, April 1997.

[KES] S. Keshav - "Congestion Control in Computer Networks". Ph.D Dissertation. Department of EECS at UC Berkeley, August 1991.

[PAX] Vern Paxson. Measurements and Analysis of End-to-End Internet Dynamics. Ph.D. dissertation, University of California, Berkeley, April 1997.

[CAR] Robert L. Carter and Mark E. Crovella. Measuring bottleneck link speed in packet-switched networks. Technical Report TR-96-006, Boston University Computer Science Department, Boston, MA, USA, March 1996.

[DOV] Constantinos Dovrolis, Parameswaran Ramanathan, David Moore What Do Packet Dispersion Techniques Measure In Proceedings of IEEE INFOCOM 2001 http://www.cs.utk.edu/~dunigan/pktprb.ps

[LAI1] Kevin Lai, Mary Baker, Nettimer: A Tool for Measuring Bottleneck Link Bandwidth, In Proceedings of the USENIX Symposium on Internet Technologies and Systems March 2001

[LAI2] Kevin Lai and Mary Baker. Measuring link bandwidth using a deterministic model of packet delay. In SIGCOMM 2000 Conference Proceedings, Stockholm, Sweden, August 28-September 01, 2000.

[DOW] Allen B. Downey. Using pathchar to estimate Internet link characteristics. In SIGCOMM '99 Conference Proceedings, pages 241–250, Cambridge, MA, USA, August 31–September 3, 1999. ACM SIGCOMM Computer Communication Review, 29(4).

[JIA] W. Jiang, T. F. Williams, Detecting and measuring Asymmetric Links in IP Network, Tech Rep, CUCS-009-99, Columbia University, 1999
http://www.cs.columbia.edu/~wenyu/papers/asym-gi99-ea.ps

[POS] J. Postel. Internet control message protocol. Request for Comments (Standard) 792, Internet Engineering Task Force, September 1981.

[BWP] http://bandwidthplace.com/speedtest/